\documentclass[9pt,twocolumn,twoside]{optica}
\setboolean{shortarticle}{false}
\setboolean{minireview}{false}

\usepackage{amsmath,amssymb,bm, float,graphicx,color,filecontents}

\title{Customizing Speckle Intensity Statistics}

\author[1]{Nicholas Bender}
\author[1]{Hasan Y{\i}lmaz}
\author[2]{Yaron Bromberg}
\author[1,*]{Hui Cao}

\affil[1]{Department of Applied Physics, Yale University, New Haven CT 06520, USA}
\affil[2]{Racah Institute of Physics, The Hebrew University of Jerusalem, Jerusalem 91904, Israel}

\affil[*]{Corresponding author: hui.cao@yale.edu}

\ociscodes{(030.0030) Coherence and statistical optics; (030.6140) Speckle; (030.6600) Statistical optics.}
\begin{abstract}

We develop a general method for customizing the intensity statistics of speckle patterns on a target plane. By judiciously modulating the phase-front of a monochromatic laser beam, we experimentally generate speckle patterns with arbitrarily-tailored intensity probability-density functions. Relative to Rayleigh speckles, our customized speckles exhibit radically different topologies yet maintain the same spatial correlation length. The customized speckles are fully developed, ergodic, and stationary: with circular non-Gaussian statistics for the complex field. Propagating away from the target plane, the customized speckles revert back to Rayleigh speckles. This work provides a versatile framework for tailoring speckle patterns with varied applications in microscopy, imaging and optical manipulation.
    
\end{abstract}

\setboolean{displaycopyright}{true}

\begin{document}
	
\maketitle
    
\section*{Introduction}

	Speckle formation is a phenomenon inherent to both classical and quantum waves. Characterized by a random granular structure, a speckle pattern arises when a coherent wave undergoes a disorder-inducing scattering process. The statistical properties of a speckle pattern are generally universal -commonly referred to as Rayleigh statistics- featuring a circular-gaussian distribution for the complex-field joint probability density function, and a negative-exponential intensity probability density function (PDF) \cite{Goodman1, GoodmanB, Dainty1, DaintyB}. Typically, non-Rayleigh speckles are classified as either under-developed (sum of a small number of scattered waves or the phase of these waves not fully randomized) or partially-coherent (sum of incoherent partial waves) \cite{Dainty2, Dainty3, Goodman2, Asakura1, Asakura2, Jakeman1, Jakeman2, Pedersen, ODonnell}. In both cases, the diversity of the intensity PDF's functional form is limited. 
	
	There has been a plethora of interest in creating speckle patterns with tailored statistics \cite{Eimerl, Goodman3, NonR1, Yaron,  NonR2, NonR3, NonR4, NonR5, Guillon1, Boyd1} due to their potential applications in structured-illumination imaging; for example: dynamic speckle illumination microscopy \cite{Dynamic2, Dynamic4}, super-resolution imaging \cite{Super_opt1,  Super_acous1},  and pseudo-thermal light sources for high-order ghost imaging \cite{image1,image2, image3}. Furthermore, a general method for customizing both the statistics and topology of laser speckle patterns would be a valuable tool for synthesizing optical potentials for cold atoms \cite{cold3}, microparticles \cite{coll1, coll2, coll3, coll4}, and active media \cite{active1, active2, active3}.

	Recently, a simple method of creating non-Rayleigh speckle patterns with a phase-only spatial light modulator (SLM) was developed \cite{Yaron}.  High-order correlations were encoded into the field by the SLM, leading to a redistribution of the light intensity among the speckle grains in the far-field. In this method, the speckle pattern could either possess an intensity PDF with a tail decaying slower or faster than a negative-exponential function. It was not known, however, if it was possible to have an intensity PDF of any functional form, such as increasing with intensity, or double-peaked at specific values.  

	In this work, we present a general method for tailoring the intensity statistics of speckle patterns by modulating the phase front of a laser beam with a SLM. Starting with a Rayleigh speckle pattern, we numerically apply a local intensity transformation to obtain a new speckle pattern which is governed by a target PDF. Subsequently this pattern is experimentally generated in the far field of the SLM, where the requisite phase modulation is determined numerically via a nonlinear optimization algorithm. Via this process, we can create speckle patterns governed by arbitrary intensity PDFs: within a predefined intensity range of interest. Such speckle patterns exhibit distinct topologies relative to Rayleigh speckles, but retaining the same spatial correlation length. A thorough study on the statistical properties reveals that the speckles are of a new kind: which has not been reported before. Despite being fully-developed, ergodic, and stationary, the joint complex-field PDFs of the speckle patterns are circular non-Gaussian; with higher-order intensity moments differing from those of Rayleigh speckles. Both intensity statistics and speckle topology evolve with beam propagation away from the target plane, wherein the speckle patterns eventually revert back to Rayleigh speckles. This work provides a versatile framework for customizing speckle patterns for varied applications in microscopy, imaging and optical manipulation.

\section*{Experimental setup and method}
\begin{figure*}[ht]
\centering
\includegraphics[width=\linewidth]{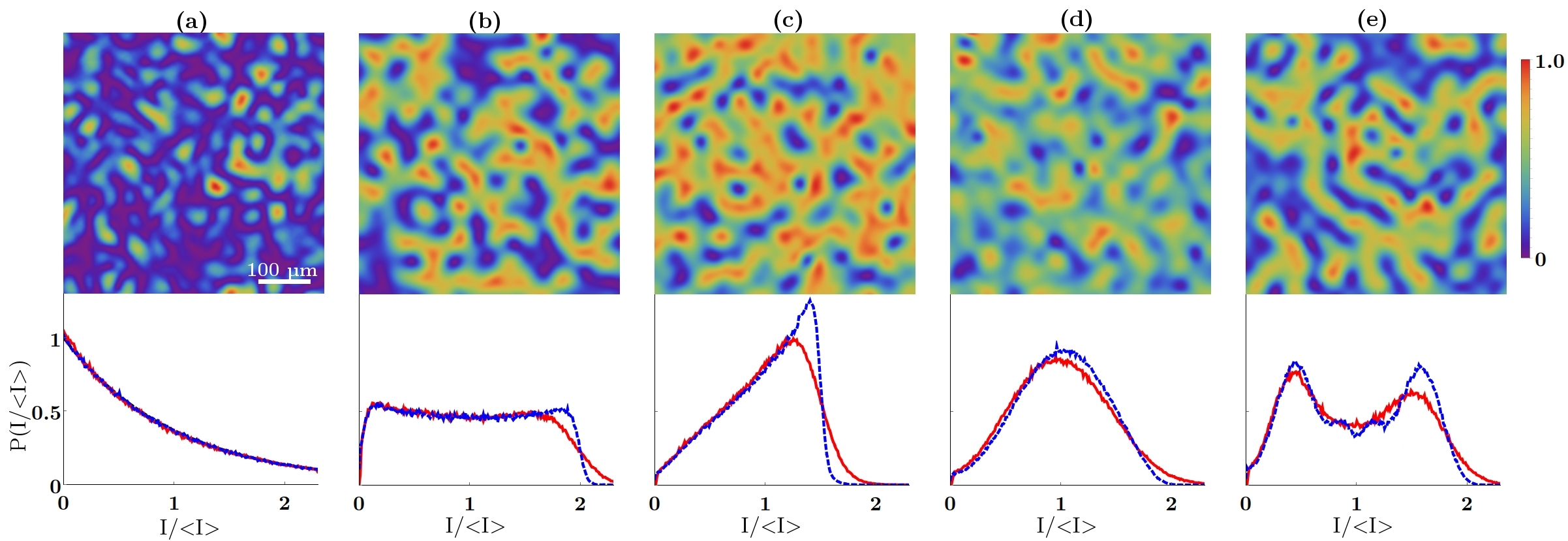}
\caption{A Rayleigh speckle pattern (a) and customized speckle patterns with distinct intensity statistics (b-e). In the top row, each pattern has the size of 504 \textmu$\mathrm{m}$ by 504 \textmu$\mathrm{m}$, and the maximum intensity is normalized to 1. The associated PDF, shown in the lower row, is uniform (b), linearly increasing (c), peaked at a non-zero intensity (d), bimodal (e), within a predefined range of intensity.  The red solid curves are experimental data, the blue dashed curves are from numerically generated target speckle patterns. Both are the result of spatial and ensemble averaging over 50 independent speckle patterns.}
\label{figure1}
\end{figure*}    
	In our experiment, a reflective, phase-only SLM (Hamamatsu LCoS X10468) is illuminated with a linearly-polarized monochromatic laser beam at the wavelength $\lambda = 642$ nm (Coherent OBIS). The laser beam is expanded and clipped by an iris to uniformly illuminate the SLM phase modulation region. The central part of the phase modulating region of the SLM is partitioned into a square array of $32 \times 32$ macro-pixels, each consisting of $16 \times 16$ pixels. The remaining illuminated pixels outside the central square diffract the laser beam away from the target plane via a phase grating. The SLM is placed at the front focal plane of a lens, $f=500$ mm, and the intensity pattern at the back focal plane is recorded by a charge-coupled device (CCD) camera (Allied Vision Prosilica GC660). The laser beam incident upon the SLM is linearly polarized and the incident angles on the camera are too small to introduce a significant polarization component in the axial direction. Thus the light waves incident on the camera can be modeled as scalar waves. To a good approximation, the field pattern on the camera chip is the Fourier transform of the field of the SLM surface. 

	Next we describe how to determine a target speckle intensity-pattern governed by an arbitrary intensity-PDF. When a Gaussian random phase pattern is displayed on the SLM, the intensity pattern recorded by the camera is a Rayleigh speckle pattern, as shown in Fig. \ref{figure1}(a). We numerically perform a local intensity transformation on a recorded Rayleigh speckle pattern which converts it into a speckle intensity-pattern governed by the desired PDF, $F(\tilde{I})$. The intensity PDF of the Rayleigh speckle pattern, $P(I)=\exp[-I/\langle I\rangle]/ \langle I\rangle]$, can be related to the target PDF $F(\tilde{I})$ by: 
    
\begin{equation}
\label{eq1}
P(I) dI = F(\tilde{I}) d\tilde{I}
\end{equation}
This relation is the starting point for determining the local intensity transformation $\tilde{I}=f(I)$, which is applied to the intensity values of a Rayleigh speckle pattern to create a new speckle pattern with the desired PDF. To solve for the specific intensity transformation associated with the PDF $F(\tilde{I})$, we write Eq. \ref{eq1} in integral form with $\langle I \rangle=1$:
\begin{equation}
\label{int}
\int_{0}^{I} e^{-I'} dI' = \int_{\tilde{I}_\mathrm{min}}^{\tilde{I}} F(\tilde{I}') d\tilde{I}'
\end{equation}
    
	Evaluating the integrals and solving for $\tilde{I}$ as a function of $I$ gives the desired local intensity transformation $\tilde{I} = f(I)$. In addition to altering the intensity PDF, such a transformation provides the freedom to regulate the maximum or minimum intensity values of the transformed pattern. We may arbitrarily set $\tilde{I}_\mathrm{max}$ or $\tilde{I}_\mathrm{min}$,  as long as the following normalization conditions hold: $\int_{\tilde{I}_\mathrm{min}}^{\tilde{I}_\mathrm{max}} F(\tilde{I}') d\tilde{I}' = 1$, and $\langle \tilde{I}\rangle = \int_{\tilde{I}_\mathrm{min}}^{\tilde{I}_\mathrm{max}} \tilde{I}' F(\tilde{I}') d\tilde{I}'=\langle I\rangle$. Such regulatory ability is useful for practical applications such as speckle illumination, where without altering the total power of illumination the maximal intensity value can be set below the damage threshold of a sample or the minimum intensity value can be set to exceed the noise floor.

	The local intensity transformation is typically nonlinear and therefore produces spatial frequency components that are higher than those in the original pattern: and therefore outside the range of spatial frequencies accessible in the experiment. Nevertheless, these components can be removed from the intensity pattern by applying a digital low-pass Fourier filter: where the allowed frequency window is a square. The high frequency cut-off of the filter is determined by the average value of the maximum spatial-frequency component present in Rayleigh speckle patterns generated by our setup. The resulting filtered-pattern, however, will have an intensity PDF $\tilde{F}(I)$ slightly deviating from the target one $F(\tilde{I})$. Such deviations can be eliminated by applying an additional intensity transformation $\tilde{I}=\tilde{f}(I)$ that is obtained from $\int_{I_\mathrm{min}}^{I} \tilde{F}(I') dI' = \int_{\tilde{I}_\mathrm{min}}^{\tilde{I}} F(\tilde{I}') d\tilde{I}'$. The process of performing a local intensity transformation, and subsequently applying a digital low-pass Fourier filter, can be iteratively repeated as the conventional Gerchberg-Saxton method \cite{GSA} until the target PDF is obtained for a speckle pattern obeying the spatial frequency restrictions. Repetition of this procedure with different initial Rayleigh speckles creates a set of independent intensity patterns which possesses the same PDFs. It is important to note, however, that certain PDFs cannot be generated in our experiment because of the finite range of spatial frequencies (See supplementary information for more detail).
	
	Having created the intensity patterns with the desired PDF, the next step is to determine the phase pattern on the SLM to generate the target pattern on the camera. Assuming the SLM consists of a $N\times N$ array of macro-pixels, a discrete Fourier transform gives a $N \times N$ array of independent elements, each representing a speckle grain. To avert the effects of aliasing and uniquely define the spatial profile measured by the camera, it is necessary to sample the speckle pattern at or above the Nyquist limit. This means every speckle grain should be sampled at least twice along each spatial axis. Thus the $N\times N$ speckles generated on the camera chip must be sampled by at least $2N\times 2N$ points. We note that the $2N \times 2N$ intensity array contains correlations between adjacent elements.  
	
	Because the phases on the SLM are transcendentally related to the intensity values on the Fourier plane, there is no closed form solution for the $N^{2}$ independent phases of the SLM macropixels to generate the $4N^{2}$ partially correlated intensity values of the target speckle pattern. Thus, we have to find the solution numerically. To facilitate the convergence to a numerical solution, we reduce the controlled area in the camera plane to the central quarter of $N^{2}$, denoted the target region, and neglect the remaining three quarters, denoted the buffer zone. Experimentally we record the speckle pattern well above the Nyquist limit with the CCD camera: $\sim 10$ camera pixels per speckle grain along each axis. Such over-sampling does not affect the solution due to spatial correlations within individual speckle grains.

	Experimentally the Fourier relation between the field reflected off the SLM and that in the camera plane is only approximate, and we characterize the precise relation by measuring the field transmission matrix (T-matrix). In addition to encapsulating the experimental imperfections induced by optical misalignment, SLM surface curvature, lens aberrations, and nonuniform laser illumination of the SLM, employing the transmission matrix provides a general formalism that can be adapted to other setups (e.g., holographic optical tweezers \cite{tweezer1}) or to tailor the speckle statistics at a different plane than the Fourier plane. In this work, the T-matrix is measured with a common path interference method akin to those in \cite{Poppoff, Yoon, Wade}. Briefly, the phase modulating region of the SLM is divided into two equal parts. We sequentially display a series of orthogonal phase patterns on one part while keeping the phase pattern on the second part fixed. Simultaneously, we record the resulting interference patterns on the camera. Subsequent to this, we exchange the role of each part and repeat the measurement. Using all the interference patterns we can construct a linear mapping between the field on the SLM and the field on the camera: the T-matrix. The measured T-matrix only slightly deviates from a discrete Fourier transform. Using the measured T-matrix, we find the requisite phase pattern on the SLM with a nonlinear-optimization algorithm \cite{nLopt1, nLopt2}. Numerically we minimize the difference between the target intensity pattern and the pattern obtained by applying the T-matrix to the SLM phase array. Starting with the SLM phase pattern that generates the original Rayleigh speckle, the algorithm converges to a solution for the SLM phase array which generates a given intensity pattern in the target region of the camera plane.
    
\section*{Customized Speckle Patterns}

	Examples of experimentally generated speckle patterns with customized intensity statistics are shown in Fig. \ref{figure1} (b-e) with $\langle I \rangle$ normalized to 1. In (b) the speckle pattern was designed to have a uniform intensity PDF, over the predefined intensity range of $I= 0$ and $2$. This example illustrates that it is possible to create speckle patterns with non-decaying PDFs in addition to confining the speckle intensities within a finite range. Taking this one step further in (c), we first make the PDF increase linearly with intensity $P(I) =I$, then have it drop rapidly to zero above the specified threshold of $I= \sqrt{2}$. To demonstrate that our method is not restricted to monotonic functions, in (d) we create a speckle pattern with a unimodal intensity PDF given by $\sin[(\frac{\pi}{2}) I]^{2}$ between $I_\mathrm{min} = 0$ and $I_\mathrm{max} = 2$. To further increase the complexity of the speckle statistics, (e) shows an example of a bimodal PDF (see the supplementary material for more information). 
    
    In all cases, the experimentally generated speckle patterns possess intensity PDFs which follow the target functional form over the intensity ranges of interest and converge to zero, quickly, outside. Small deviations between the experimental PDFs and the target ones are caused by error in the T-matrix measurement due to experimental noise and temporal decorrelations. We modeled these effects and numerically reproduced the deviations (see the supplementary material for a full description of our model). Our model describes why the deviations are stronger at higher intensity values or where the PDF varies rapidly with intensity.
	
	Additionally Fig. \ref{figure1} illustrates how the topology of the customized speckle patterns changes in accordance with the PDF. The spatial intensity profile of a Rayleigh speckle pattern in (a) can be characterized as a random interconnected web of dark channels surrounding bright islands. Conversely for speckles with a linearly increasing PDF in (c), the spatial intensity profile is an interwoven web of bright channels with randomly dispersed dark islands. Similarly the spatial structure of speckles with a bimodal PDF in (e) consists of interlaced bright and dim channels. The topological changes in the customized speckles result from the local intensity transformation and digital low-pass Fourier filtering (see the supplementary material for details). The continuous network of high intensity in the customized speckle pattern, which is absent in the Rayleigh speckle pattern, will be useful for controlling the transport of trapped atoms or microparticles in optical potentials. 

\section*{Statistical Properties of Customized Speckle Patterns}

	In this section, we analyze the statistical properties of the customized speckle patterns to illustrate their stark difference relative to previously studied speckles. 
    
\begin{figure}[htb]
\centering
\includegraphics[width=\linewidth]{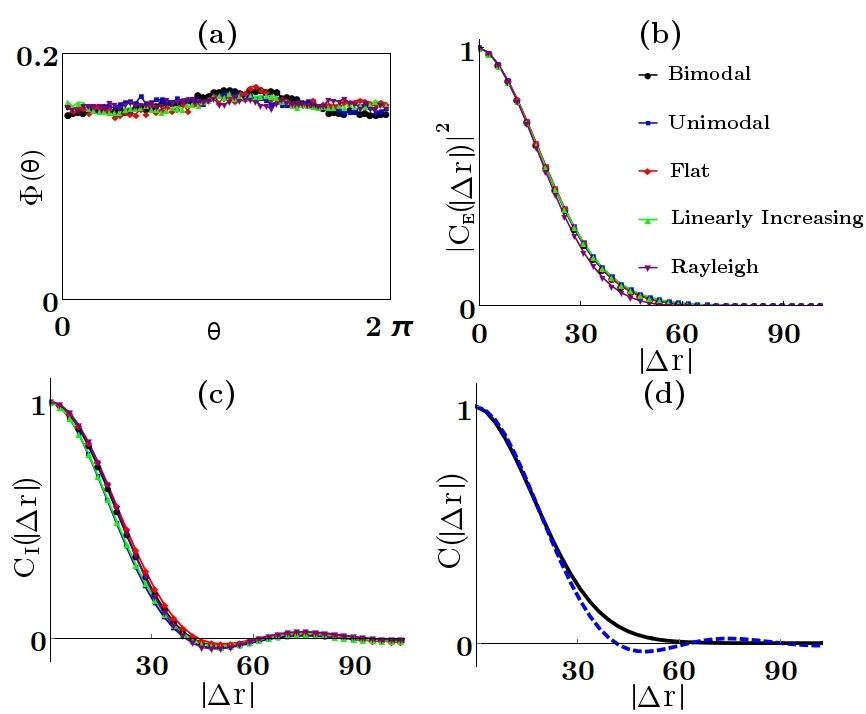}
\caption{Characteristics of customized speckle patterns. (a) The phase histogram of the speckle fields demonstrates that they are all fully developed speckles. (b) The spatial field correlation function $|C_{E}(|\Delta {\bf r}|)|^2$ and (c) the spatial intensity correlation function $C_{I}(|\Delta {\bf r}|)$, for the customized speckles, remain the same as in Rayleigh speckles. In (a-c), the four customized speckle patterns have constant (black), linear increasing (red), unimodal (blue), bimodal (green) PDFs, and the purple is for Rayleigh speckles. (d) Comparing $|C_{E}(|\Delta {\bf r}|)|^2$ (black curve) to $C_{I}(|\Delta {\bf r}|)$ (blue dashed curve), both averaged over the 5 curves in (b) and (c) respectively, to confirm they have the same correlation width.}
\label{figure2}
\end{figure}

    We start by verifying that the speckle patterns are fully developed and the phase distribution $\Phi(\theta)$ of the generated speckle fields is uniform over a range of $2 \pi$. To find $\Phi(\theta)$, we use the measured transmission matrix and the SLM phase patterns to recover the fields associated with the intensity patterns recorded by the CCD camera. Figure \ref{figure2}(a) plots $\Phi(\theta)$ in the target region for the four customized PDFs shown in Figure \ref{figure1}(b-e), in addition to the case of a Rayleigh PDF. All cases have nearly constant values over $[0, 2 \pi]$, thus our speckle patterns are fully developed. This property differentiates our speckle patterns with a unimodal PDF, shown in Figure \ref{figure1} (d), from partially developed speckle patterns which possess a similar intensity PDF \cite{GoodmanB}. 
    
	Next, we check whether additional spatial correlations are introduced into the customized speckle patterns, relative to Rayleigh speckles. To this end, we calculate the 2D spatial correlation function of the speckle field:  $C_{E}(\Delta {\bf r}) = \langle \tilde{E}({\bf r}) \tilde{E}^*({\bf r} + \Delta {\bf r})\rangle / \langle \tilde{I} \rangle $. As shown in Fig. \ref{figure2}(b), the customized speckles have a similar field correlation function as a Rayleigh speckle pattern. This means the way we tailor the speckle statistics does not affect the spatial field-correlation function. Furthermore, the 2D spatial intensity-correlation function, $C_{I}(\Delta {\bf r}) = (\langle \tilde{I}({\bf r}) \tilde{I}({\bf r} + \Delta {\bf r})\rangle-\langle \tilde{I} \rangle^{2})/(\langle \tilde{I}^{2} \rangle-\langle \tilde{I}\rangle ^{2})$, plotted in Fig. \ref{figure2}(c) for the four customized speckles, has the same width as a Rayleigh speckle pattern. Hence, we can manipulate the speckle intensity PDF without altering the spatial correlation length.
    
\begin{figure}[htb]
\centering
\includegraphics[width=\linewidth]{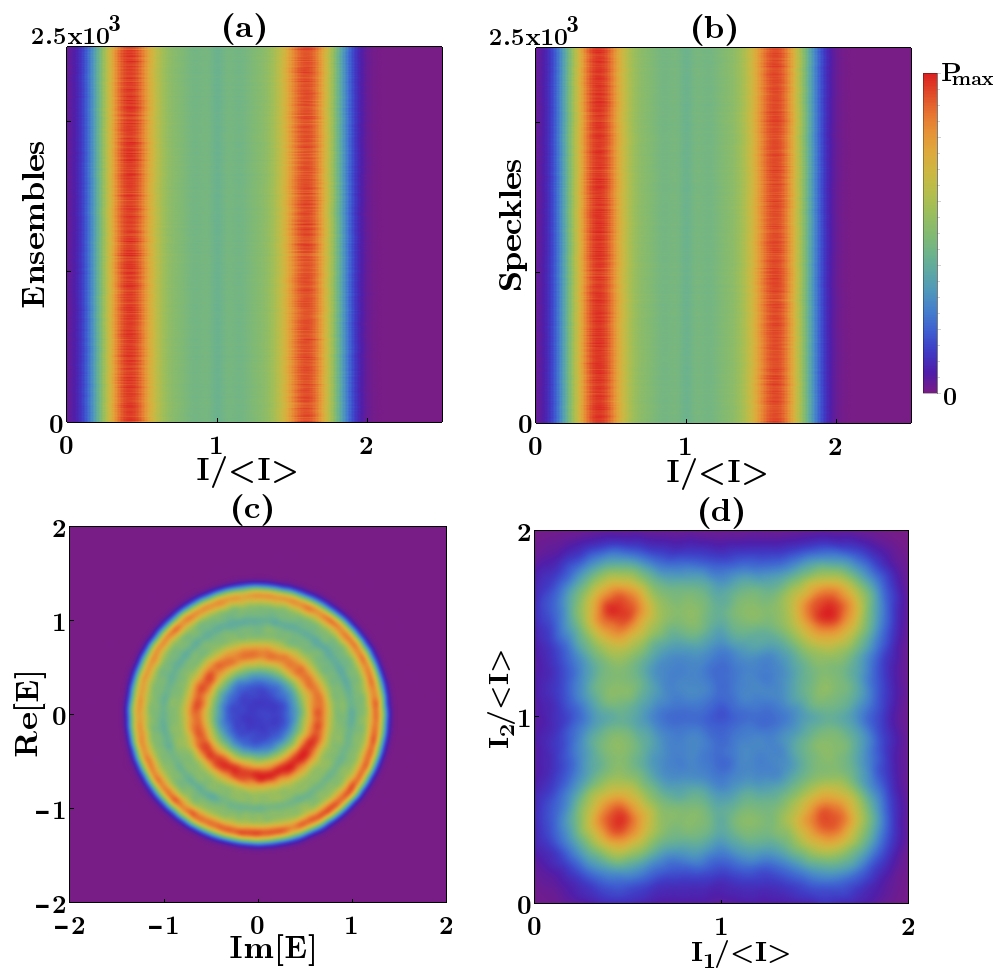}
\caption{The statistical properties of the customized speckle patterns with a bimodal intensity PDF. (a) The spatially averaged intensity PDF $P_S(I)$ of 2500 speckle patterns verifies that the speckle patterns are stationary. (b) The ensemble-averaged intensity PDF $P_E(I)$ -within the area of a single speckle grain- for each spatial position in a pattern confirms the speckles are ergodic. (c) The complex-amplitude joint-PDF for the speckle field $P({\rm Re}[E], {\rm Im}[E])$ displays circular non-Gaussian statistics. (d) The joint PDF for two intensities $P(I_1, I_2)$ at spatial positions separated by approximately one speckle grain size (twice of the spatial intensity correlation width) are uncorrelated. The results in (c,d) are obtained from averaging over space and 100 speckle patterns, a digital low-pass Fourier filter is applied to remove noise.}
\label{figure3}
\end{figure}

	 Similar to the Rayleigh speckles, the customized speckle patterns have the same width for $\langle |C_{E}(\Delta {\bf r})|^2 \rangle$ and $\langle C_{I}(\Delta {\bf r}) \rangle$, as shown in Fig. \ref{figure2}(d). Relative to $\langle |C_{E}(\Delta {\bf r})|^2 \rangle$, however, $\langle C_{I}(\Delta {\bf r}) \rangle$ exhibits small oscillations on the tail. These are attributed to the low-pass Fourier filtering of the intensity pattern, which we use to remove the high spatial frequency components introduced during the nonlinear transformation of a Rayleigh speckle pattern. For confirmation of this, we applied the digital low-pass Fourier filter to Rayleigh speckle patterns and the same oscillations appeared in the spatial intensity correlation function, shown in Fig. \ref{figure2}(c). 
    
\begin{figure*}[ht]
\centering
\includegraphics[width=14cm]{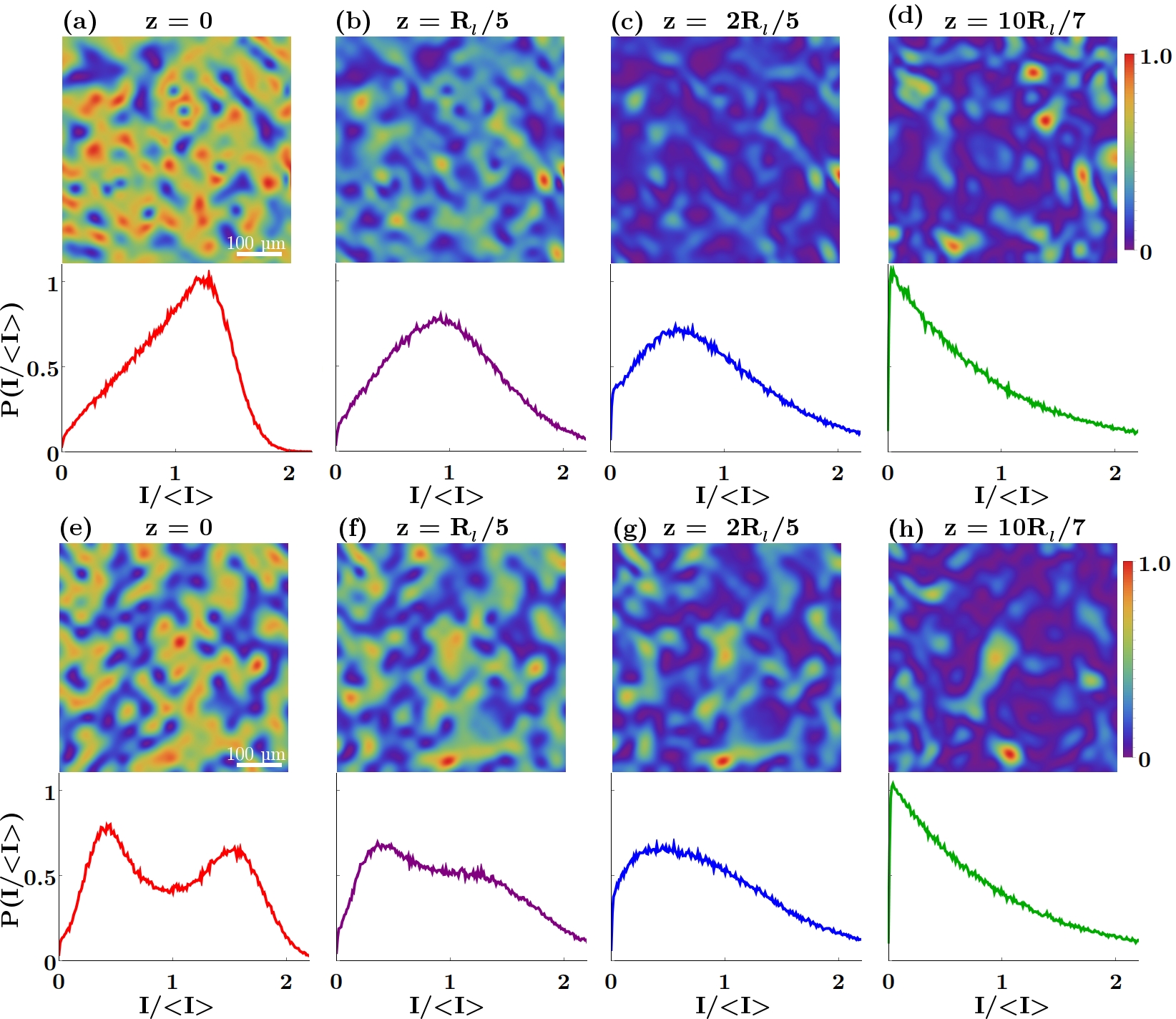}
\caption{Evolution of customized speckle patterns upon axial propagation. The intensity PDF at the Fourier plane of the SLM ($z=0$) is linearly increasing (a) and bimodal (e). The distance from the Fourier plane is ${R_{l}}/{5}$ (b, f), $(2/5) R_{l}$ (c, g), and $(10/7) R_{l}$ (d, h).}
\label{figure4}
\end{figure*}
       
	Now we investigate whether the generation of tailored speckle patterns can be described statistically as a stationary and ergodic process. For a target intensity PDF, we numerically create 2500 speckle patterns consisting of 2500 speckle grains each. Fig. \ref{figure2}(a,b) shows the results for the bimodal PDF. In (a), the intensity PDF obtained for each of the 2500 ensembles is invariant as a function of ensembles, indicating the speckle patterns are stationary. In (b), the ensemble-averaged intensity PDF for individual spatial positions in the speckle patterns is also invariant as a function of spatial position and statistically identical to (a), demonstrating the ergodicity of the speckle patterns.  

Figure \ref{figure3}(c) shows the joint complex-field PDF of the bimodal speckles is circular non-Gaussian, in contrast to the circular Gaussian PDF of Rayleigh speckles. Circularity reflects the fact that the amplitude and phase of the speckle field $E$ -at a single point in space- are uncorrelated \cite{GoodmanB,DaintyB}. Figure \ref{figure3}(d) shows the joint PDF $P(I_{1}, I_{2})$ for two speckle intensities at locations separated by more than the average speckle grain size. Because $P(I_{1}, I_{2}) = P(I_{1}) P(I_{2})$ (see the supplementary material), the two intensities are statistically independent, which is consistent with the spatial correlation length of speckle intensity. The other types of customized speckle patterns display similar characteristic, and the results are presented in the supplementary material.

\begin{table}[htbp]
\centering
\caption{\bf Intensity moments of speckle patterns with different intensity PDFs}
\begin{tabular}{ccccccc}
\hline
PDF & $\langle I \rangle$ & $\langle I^{2} \rangle$ & $\langle I^{3} \rangle$ & $\langle I^{4} \rangle$ & $\langle I^{5} \rangle$ & $\langle I^{6} \rangle$\\
\hline
Negative Exponential & 1.00 & 2.00 & 6.00 & 24.0 & 120 & 720\\
Constant & 1.00 & 1.35 & 2.06 & 3.39 & 5.87 & 10.51\\
Linearly Increasing & 1.00 & 1.16 & 1.45 & 1.92 & 2.64 & 3.77\\
Unimodal &  1.00 & 1.18 & 1.55 & 2.22 & 3.40 & 5.50\\
Bimodal & 1.00 & 1.29 & 1.9 & 2.99 & 4.93 & 8.42\\
\hline
\end{tabular}
\label{table}
\end{table} 

	The non-Gaussian statistics of the tailored speckle patterns also emerge in their high-order intensity moments, $\langle I^{n} \rangle = \int_{0}^{\infty} I^{n} P(I) dI$, which differ from those of Rayleigh speckles: shown in Table \ref{table}. For the case of Rayleigh speckles generated by a Gaussian-random process, the high-order moments are related by $ \langle I^{n} \rangle = n! \langle I \rangle^n$ \cite{Dainty1}. For the customized speckles, $\langle I^{n} \rangle$ deviates from $ n! \langle I \rangle^n$, because of high-order correlations among the partial waves that generate these patterns \cite{Yaron}.
    
\section*{Axial Propagation}
  
	Finally, we study how the tailored speckle patterns evolve as they propagate axially. Our method gives the target PDF for speckle patterns on the Fourier plane of the SLM. Outside of the Fourier plane, however, the intensity statistics and topology may change. In the case of a Rayleigh speckle pattern, the spatial pattern changes upon propagation while the intensity PDF remains a negative exponential. We define $R_{l}$ as the axial correlation length of the intensity pattern, which corresponds to the Rayleigh range and gives the longitudinal length of a single speckle grain.
    
	The top row of Fig. \ref{figure4} shows the axial evolution of speckles that have a linearly increasing PDF at the Fourier plane $z=0$ in (a). As the speckle pattern propagates to $z=R_{l}/5$, the PDF becomes bell-shaped in (b). With further propagation, the maximum of the PDF migrates to a smaller intensity value, as shown in (c) for $z=(2/5)R_{l}$, until it reaches $I=0$. The speckles revert to Rayleigh statistics at $z\approx R_{l}$, beyond which the PDF maintains a negative exponential, as shown in (d) for $z=(10/7) R_{l}$. The topology of the speckle pattern evolves together with the intensity PDF: the interconnected web of bright channels first attenuates upon propagation, then breaks into islands with dark channels forming. 
	
	In Fig. \ref{figure4} (e-f), we show the axial evolution of a speckle pattern with a bimodal PDF at $z=0$ in (e). As the pattern propagates to $z={R_{l}}/{5}$ in (f), the peaks are asymmetrically eroded, with the high intensity peak diminishing first. Further propagation to $z=(2/5)R_{l}$ results in a unimodal PDF in (g). Once the axial distance $z$ exceeds $R_{l}$, the speckles return to Rayleigh statistics, as shown in (h) for $z=(10/7) R_{l}$. A corresponding change of speckle topology is seen: the bright channels disappear first, then the dim channels fracture, while neighboring dark islands merge to form channels. Therefore, axial propagation, within the range of $R_l$, alters the intensity PDF functional form and the speckle topology. 
    
\section*{Conclusion}
    
	In conclusion, we have presented a general method for customizing speckle intensity statistics using a phase-only SLM. The generated speckle patterns possess radically different intensity PDFs and topology relative to Rayleigh speckles. However, they are fully developed speckles which maintain the basic characteristics of stationarity and ergodicity. Their unusual statistical properties engender a new type of speckle pattern with non-Gaussian statistics. Our method is versatile and compatible with a broad range of optical setups. Given the plethora of potential applications, it paves the way for new directions in both fundamental research (many-body physics in random optical potentials with tailored statistics) and applied research (speckle-illumination-based imaging and speckle optical tweezers).
    
\subsection*{Acknowledgment.} We thank Chia Wei Hsu and Owen Miller for useful discussions. 
\subsection*{Funding.} This work is supported partly by the MURI grant no. N00014-13-1-0649 from the US Office of Naval Research. 
	
\bigskip \noindent See \href{link}{Supplement} for supporting content.
\bibliographyfullrefs{sample}

\newpage

\part*{Supplementary Material}

    \section*{Constraints on intensity PDF}
    
\begin{figure}[ht]
    	\centering
    	\includegraphics[width=6 cm]{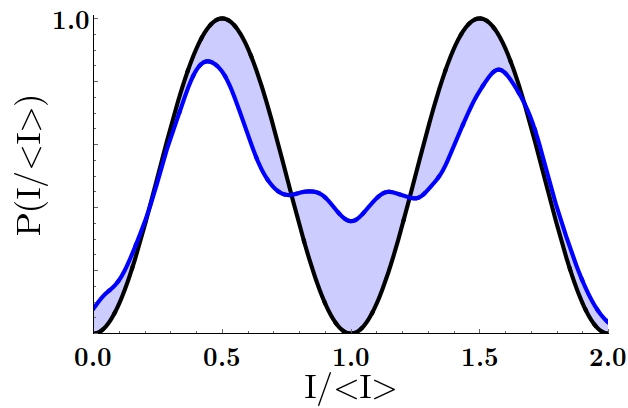}
    	\caption{The customized speckle patterns' bimodal intensity PDF. The black curve is the original desired PDF, and the blue curve is the numerically obtained one. Their difference (shaded area) is due to the limited range of the spatial frequencies in the speckle pattern.}
    	\label{figure5S}
    \end{figure}

    In the main text we mention that certain PDFs cannot be generated in our experiment because of the finite range of spatial frequencies. One example is the bimodal intensity PDF shown in Fig. 1(e) of the main text. The original functional form used to generate this is $\sin^{2}(\pi I)\, H[2-I]$, where $P(I/<I> = 1) = 0$. However, the target PDF cannot vanish or have a discontinuity at any intensity value in between the minimal and maximal intensities due to finite range of the spatial frequencies on the speckle pattern. Consequently, the iterative transformation algorithm converges to the PDF plotted in blue in Fig. \ref{figure5S}, which differs from the original one plotted in black. The blue shading in this figure highlights the regions where deviations occur, primarily centered around $I/<I>=1$.

	\section*{Experimental error of speckle customization}
	
\begin{figure}[ht]
		\centering
		\includegraphics[width= 6cm]{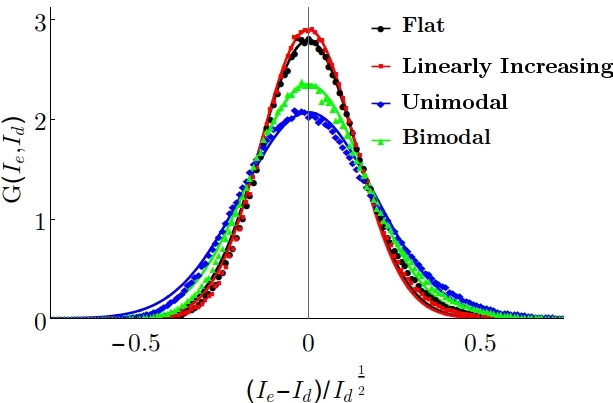}
		\caption{The statistical distribution of $(I_{e}-I_{d})/\sqrt{I_{d}}$, extracted from the experimental data (symbols), is fit well by the Gaussian distribution $G(I_{e},I_{d})$ in Eq. \ref{gaus} (lines). The fitting parameter is given by $a = 0.020$ for the uniform PDF (black),  $a = 0.019$ for the linearly increasing PDF (red), $a=0.037$ for the PDF with a single peak (blue) and  $a=0.029$ for the bimodal PDF (green).}
		\label{figure6S}
   \end{figure}

	In this section we perform a theoretical assessment of the experimental error to understand its effect on the intensity PDF of the customized speckle patterns. Due to measurement noise and temporal decorrelation of the experimental setup (e.g., due to ambient temperature drift), the measured T-matrix $\mathbf{T_{m}}$ differs from the actual T-matrix  $\mathbf{T_{a}}$ that produces the measured speckle pattern. Their difference is:
	\begin{equation}
	\Delta\mathbf{T} = \mathbf{T_{m}} - \mathbf{T_{a}}
	\end{equation}
	$\mathbf{T_{m}}$ is used in the numerical optimization to obtain the SLM field ${\vec{\Psi}}_{s}$ to create the target speckle intensity pattern
	\begin{equation}
	{\vec{I}}_{d}= |\mathbf{T_{m}} \, {\vec{\Psi}}_{s}|^{2}
	\end{equation}
	The measured speckle intensity pattern is 
	\begin{equation}
	{\vec{I}}_{e}= |\mathbf{T_{a}} \, {\vec{\Psi}}_{s}|^{2} 
	\end{equation}
	The difference between the two intensities, to the first order in $\Delta\mathbf{T}$, is
	\begin{equation}
	{\vec{I}}_{e}-{\vec{I}}_{d} = 2 \Re\big{[}(\mathbf{T}_{m} \, {\vec{\Psi}}_{s})(\Delta\mathbf{T} \, {\vec{\Psi}}_{s})^{*}\big{]} =  2 \Re\big{[}\sqrt{{\vec{I}}_{d}} \, e^{i {\vec{\theta}}_{d}}(\Delta\mathbf{T} \, {\vec{\Psi}}_{s})^{*}\big{]} \, ,
	\end{equation}
	where $\mathbf{T}_{m} \, {\vec{\Psi}}_{s} = \sqrt{{\vec{I}}_{d}} \, e^{i{\vec{\theta}}_{d}}$. Assuming the scalar elements of $\Delta\mathbf{T}$ are uncorrelated with those in ${\vec{\Psi}}_{s}$, then the scalar quantities of $(\vec{I}_{e}-\vec{I}_{d})/\sqrt{\vec{I}_{d}} = 2 \Re\big{[}(\Delta\mathbf{T} \, \vec{\Psi}_{s})^{*}\big{]}$ will obey Gaussian statistics, as verified experimentally in Fig. \ref{figure6S}. The statistical distribution for $(I_{e}-I_{d})/\sqrt{I_{d}}$ is:
	\begin{equation}\label{gaus}
	G(I_{e},I_{d})= A \exp\big{[}\frac{-(I_{e}-I_{d})^{2}}{2a  I_{d}}\big{]}
	\end{equation}
	where $a$ is a coefficient that quantifies the experimental error, and $A$ is the normalization constant given by 
	$\int_{0}^{\infty}G(I_{e},I_{d})dI_{d}=1$.    

	Therefore, the normalized expression for $G(I_{e},I_{d})$ is 
	\begin{equation}\label{dist}
	G(I_{e},I_{d})=\frac{\exp\big{[}\frac{-(I_{e})^{2}-(I_{d})^{2}}{2a I_{d}}\big{]}}{2 I_{e} K_{1}[I_{e}/a]}
	\end{equation}
	where $K_{1}$ is the Bessel function of the first kind.

	The probability density function of the measured speckle intensity $\bar{F}(I_{e})$ is equal to the target distribution $F(I_{d})$ convolved  with the error function $G(I_{e},I_{d})$ in Eq.(\ref{dist}):
	\begin{equation}\label{error}
	\bar{F}(I_{e})=\int_{0}^{\infty} F(I_{d}) G(I_{e},I_{d}) dI_{d}
	\end{equation}
 
    	\begin{figure}[ht]
		\centering
		\includegraphics[width=\linewidth]{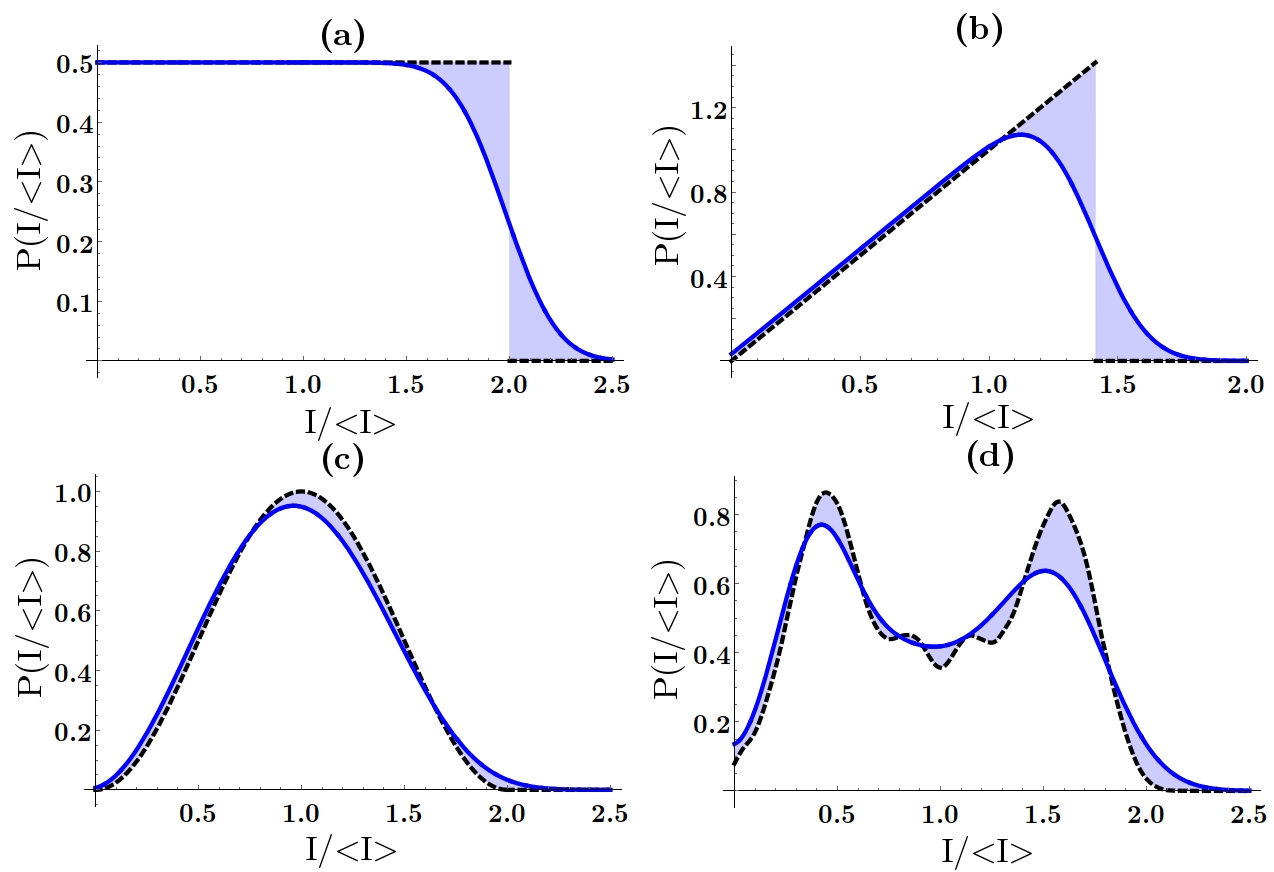}
		\caption{the deviation (shaded area) of the intensity PDF (blue solid line) from the target one (black dashed line) is reproduced numerically by Eq.  \ref{error}. (a-d) for the four PDFs shown in Fig. 1 of the main text.}
		\label{figure7S}
	\end{figure}
    
	The convolution corresponds to an averaging of $F(I_{d})$ over adjacent values of $I_d$. Consequently, the measured PDF $\bar{F}(I_{e})$ displays more discrepancy in the region where $F(I_d)$ changes rapidly. Since $G(I_{e},I_{d})$ is wider at larger $I_e$, the averaging effect is stronger, leading to a larger error at higher intensity. These effects are confirmed in Fig. \ref{figure7S}, where we plot Eq.(\ref{error}) for the four PDFs shown in Fig. 1 of the main text. For the uniform and linear increasing PDFs in (a) and (b), the abrupt drop in $F(I_{d})$ at the upper boundary of the intensity range is smoothed out in $\bar{F}(I_{e})$. In (c), the single-peaked PDF has relatively small error, although the deviation from the target PDF is clearly larger at higher intensity. For the bimodal PDF in (d), the peak at larger intensity is suppressed more due to stronger averaging effect, and the fine features around the dip in between the two peaks are removed by averaging.

\section*{Topological change of speckle patterns}

\begin{figure*}[ht]
	\centering
	\includegraphics[width=14 cm]{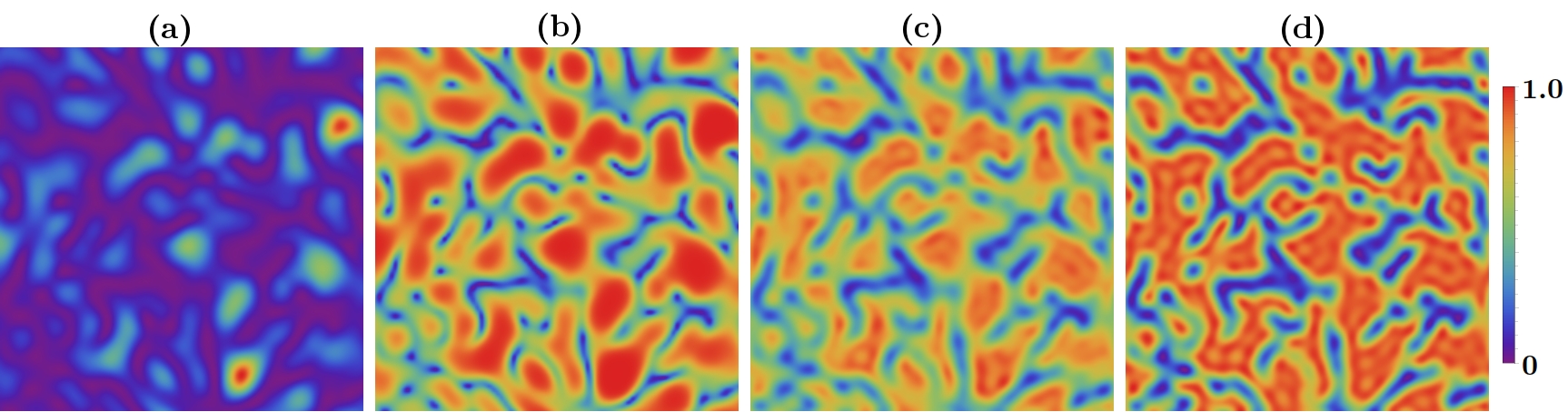}
	\caption{The topological change of a customized speckle pattern. A Rayleigh speckle pattern in (a) is transformed to the speckle pattern in (b) via a local intensity transformation to have a linearly increasing PDF. Application of a digital low-pass Fourier filter results in the pattern in (c). Multiple iterations of intensity transformations and filtering result in the final pattern in (d) which obeys the desired intensity PDF and spatial frequency constraints.}
	\label{figure8S}
\end{figure*}

As shown in Fig. 1 of the main text, the topology of the customized speckle patterns is distinct from that of Rayleigh speckle. The change in speckle topology is caused by a combination of the local intensity transformation and the digital low-pass Fourier filtering. One example is given in Fig. \ref{figure8S}. The original Rayleigh speckle pattern in (a) has a maximal probability-density of vanishing intensity, leading to the dark channels surrounding bright islands in the spatial profile of the speckle pattern. Application of a local intensity transformation to make the PDF increase linearly with intensity and then rapidly converge to zero above a threshold results in the speckle pattern shown in (b). Due to the enhanced probability-density of high intensity and reduced probability-density of low intensity, the bright grains are enlarged while the dark channels are narrowed. The application of a digital low-pass Fourier filter severs the narrow dark lines in between the bright grains. After iterating the process of a local intensity transformation followed by a low-pass filter, neighboring bright grains are merged to form channels that encompass dark islands.   

\section*{Statistical properties of customized speckles}

In the main text we show the statistical properties of the speckle patterns with a bimodal intensity PDF. In this section, we present the results for other types of customized speckles shown in Fig. 1 of the main text.

\begin{figure}[H]
		\centering
		\includegraphics[width=\linewidth]{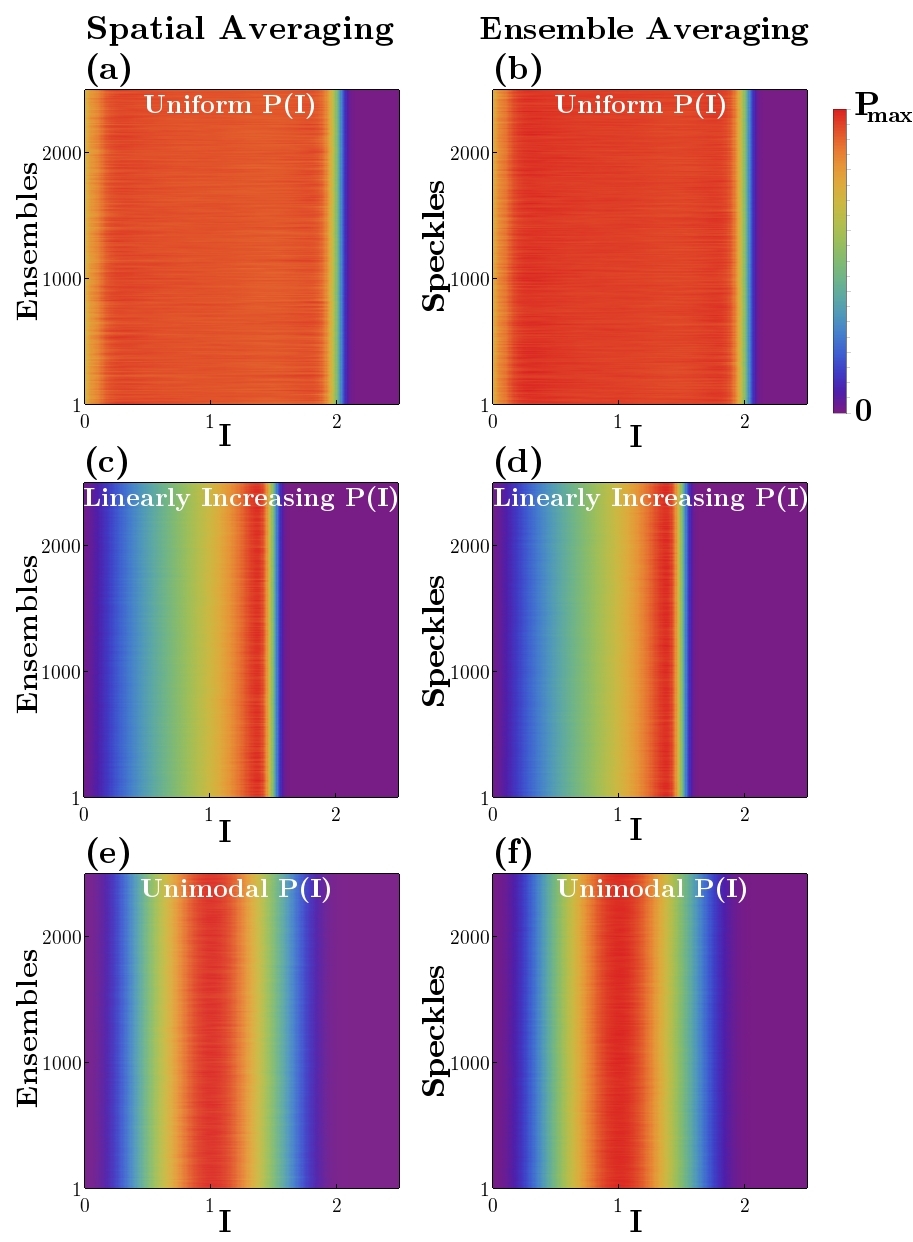}
		\caption{Left column (a,c,e): the spatially averaged intensity PDF of 2500 individual speckle patterns customized to have the same intensity PDF. Right column (b,d,f): the ensemble-averaged intensity PDF -within the area of a single speckle grain- for each spatial position in the same set of speckle patterns. The speckle intensity PDF is uniform in (a,b), linearly increasing in (c,d), and unimodal in (e,d). All the customized speckle patterns are stationary and ergodic. We set $\langle I \rangle=1$.}
		\label{figure1S}
	\end{figure}

In Figure \ref{figure1S}, we demonstrate that the generation of customized speckle patterns can be described by a stationary and ergodic process, for the uniform intensity PDF in (a,b), the linearly increasing intensity PDF in (c,d), and the unimodal intensity PDF in (e,d). Numerically we generate 2500 target speckle intensity patterns, each consisting of 2500 speckle grains, for each of the three customized intensity PDFs. The left column is the plot of spatial-averaged intensity PDF for each of the 2500 ensembles. For comparison, the right column is the plot of ensemble-averaged intensity PDFs for each spatial position in the speckle patterns. Because the three intensity PDFs in the left column are invariant with respect to different ensembles, the speckles are stationary. Because the three intensity PDFs in the right column are invariant with respect to different spatial positions and they are statistically equivalent to the ones in the left column, the speckles are ergodic. 

\newpage

\begin{figure}[H]
		\centering
		\includegraphics[width=\linewidth]{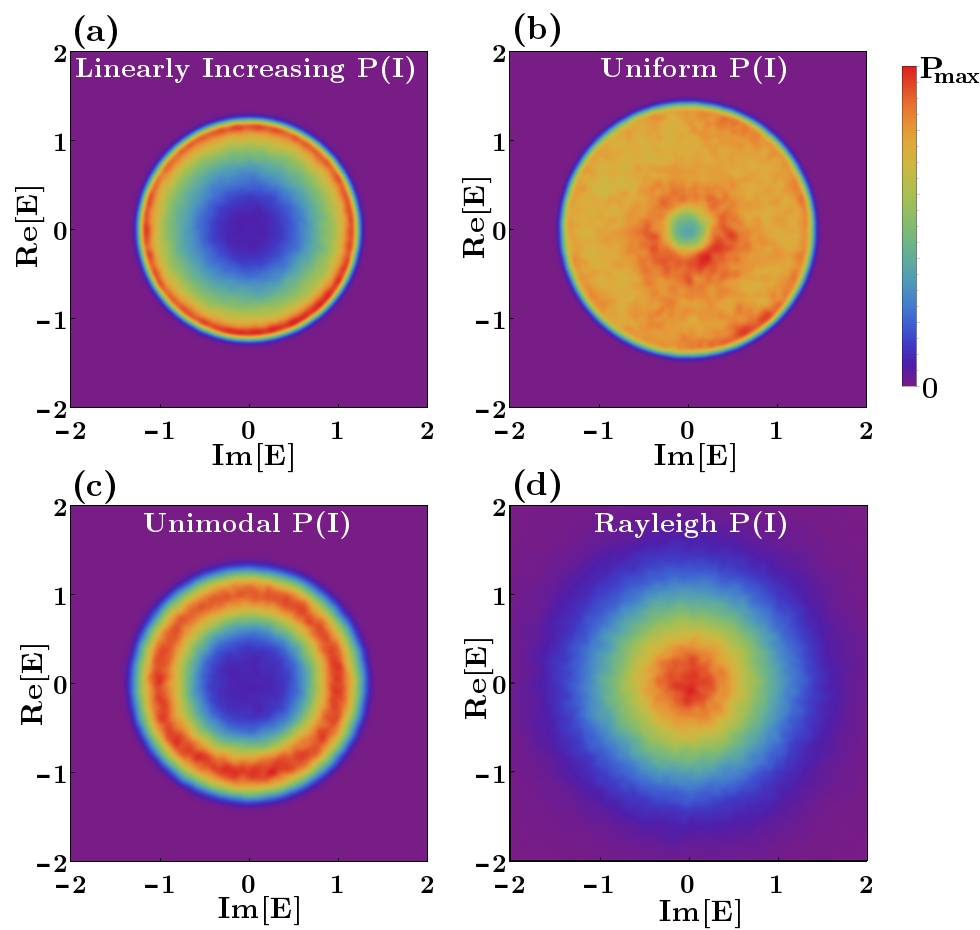}
		\caption{The complex-amplitude joint-PDF $P({\rm Re}[E], {\rm Im}[E])$ for the speckle field $E$. The corresponding intensity PDF increases linearly in (a), is uniform in (b), is unimodal in (c), and decays exponentially in (d). While the Rayleigh speckle in (d) is circular-Gaussian, the customized speckles in (a-c) are circular non-Gaussian.}
		\label{figure2S}
	\end{figure}

Figure \ref{figure2S} demonstrates that the complex-fields of the customized speckle patterns possess circular non-Gaussian statistics. The  joint complex-field PDF $P({\rm Re}[E], {\rm Im}[E])$ for the speckle field $E$ is plotted for the case of a linearly increasing intensity PDF in (a), a uniform intensity PDF in (b), a unimodal intensity PDF in (c). For reference, (d) shows the circular-Gaussian PDF of Rayleigh speckles. We average both spatially and over 100 speckle patterns for each figure, and apply a low-pass Fourier filter to remove the noise. The circularity indicates not only that the phase of speckles' field is fully randomized and distributed uniformly across $[0, 2 \pi]$, but also that the phase and amplitude of the speckles' field at any position are statistically uncorrelated.     

\newpage

\begin{figure}[H]
		\centering
		\includegraphics[width=\linewidth]{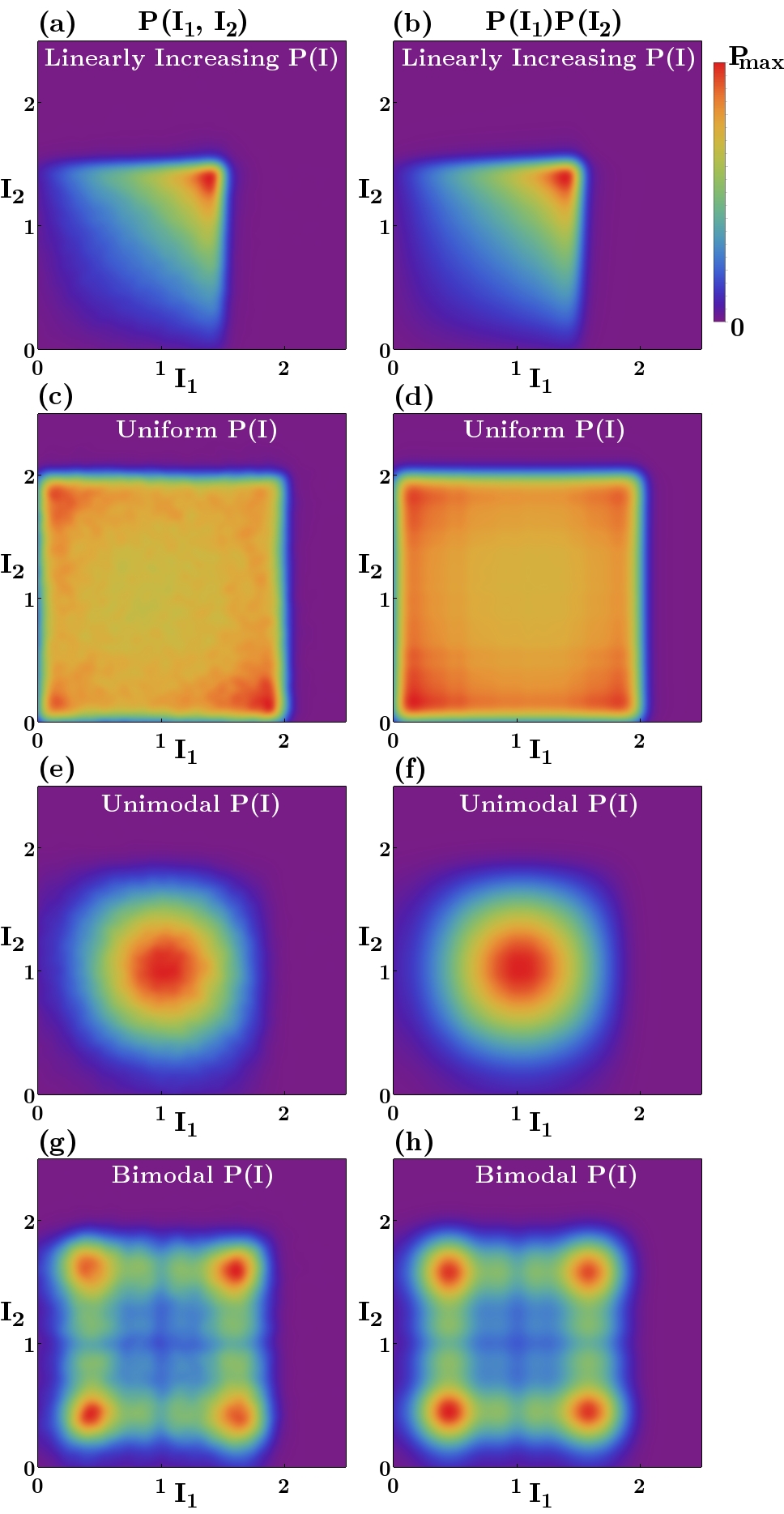}
		\caption{Joint PDFs $P(I_{1}, I_{2})$ for two intensities $I_1$ and $I_2$ at spatial positions separated by approximately one speckle grain size (left column) and the product $P(I_{1})P(I_{2})$ (right column) for speckles customized to have a linearly increasing intensity PDF in (a,b), a uniform intensity PDF in (c,d), unimodal intensity PDF in (e,f), and a bimodal intensity PDF in (g,h). Since $P(I_{1}, I_{2})$ is equal to $P(I_{1})P(I_{2})$, the two intensities $I_1$ and $I_2$ are statistically independent. We set $\langle I \rangle=1$.}
		\label{figure3S}
	\end{figure}

Figure \ref{figure3S} shows that the joint PDFs for two intensities at spatial positions separated by approximately one speckle grain size are independent for the customized speckle patterns. The left column is the plot of $P(I_{1}, I_{2})$ and the right column $P(I_{1})P(I_{2})$ for the intensity PDF that is linear increasing in (a,b),  uniform in (c,d), unimodal in (e,f), and bimodal in (g,h). PDF are obtained by averaging over space and 100 speckle patterns. The average speckle grain size is given by twice the width of the spatial intensity correlation function, and is  $\approx  60 \mu$m. Because $P(I_{1}, I_{2})$ is equal to $ P(I_{1}) P(I_{2})$ in all cases, the two intensities are statistically independent, which is consistent with the spatial intensity correlation length. Additionally, the phases of two speckle fields at spatial positions separated by more than one speckle grain size are independent for all customized speckle patterns, consistent with the spatial field correlation length.   
	
\end{document}